\documentclass[fleqn,usenatbib]{mnras}

\usepackage{newtxtext,newtxmath}

\usepackage[T1]{fontenc}

\DeclareRobustCommand{\VAN}[3]{#2}
\let\VANthebibliography\thebibliography
\def\thebibliography{\DeclareRobustCommand{\VAN}[3]{##3}\VANthebibliography}

\usepackage{graphicx}	
\usepackage{amsmath}	

\title[Spectroscopic confirmation of an AGN]{High energy gamma-ray sources in the VVV survey - III. Spectroscopic confirmation of an AGN at low Galactic latitudes}

\author[E. O. Schmidt et al.]{
E. O. Schmidt,$^{1,2}$\thanks{E-mail: eduardo.schmidt@unc.edu.ar}
L. D. Baravalle,$^{2,1}$
A. Pichel,$^{3}$
D. Mast,$^{1}$
M. V. Alonso,$^{2,1}$
J. Diaz Tello,$^{4}$
\newauthor
L. H. García,$^{1}$
V. D. Ivanov,$^{5}$ 
D. Minniti,$^{6,7}$
N. Masetti,$^{8,6}$
L. Donoso,$^{9}$
and 
R. Zelada Bacigalupo$^{10}$
\\
$^{1}$Observatorio Astron\'omico de C\'ordoba, Universidad Nacional de C\'ordoba, Laprida 854, X5000BGR, C\'ordoba, Argentina\\
$^{2}$Instituto de Astronom\'{\i}a Te\'orica y Experimental, (IATE-CONICET), Laprida 854, X5000BGR, C\'ordoba, Argentina\\
$^{3}$Instituto de Astronom\'ia y F\'isica del Espacio, CONICET–UBA, Av. Int. Guiraldes 2620, C1428BNB CABA, Argentina\\
$^{4}$Preuniversitario UC, Pontificia Universidad Católica de Chile, Chile\\
$^{5}$European Southern Observatory, Karl Schwarzschildstr 2, 85748
Garching bei München, Germany\\
$^{6}$ Instituto de Astrofísica, Dep. de Ciencias Físicas, Facultad de Ciencias Exactas, Universidad Andres Bello, Av. Fernández Concha 700, Santiago, Chile\\
$^{7}$Vatican Observatory, V00120 Vatican City State, Italy\\
$^{8}$INAF - Osservatorio di Astrofisica e Scienza dello Spazio, via Piero Gobetti 101, I-40129 Bologna, Italy\\
$^{9}$ Facultad de Ciencias Exactas, Físicas y Naturales, UNSJ, San Juan, Argentina\\
$^{10}$North Optics Instrumentos Científicos, La Serena, Chile 
}

\date{Accepted XXX. Received YYY; in original form ZZZ}

\pubyear{\the\year{}}

\begin{document}
\label{firstpage}
\pagerange{\pageref{firstpage}--\pageref{lastpage}}
\maketitle

\begin{abstract}
We aim to spectroscopically confirm the nature of VVV-J181258.71-314346.7, a candidate counterpart to the unassociated gamma-ray source 4FGLJ1812.8-3144. This object was selected based on its near-infrared photometric properties and moderate variability, as part of a broader effort to identify active galactic nuclei (AGN) behind the Galactic bulge and disc.
We obtained near-infrared spectra using the Flamingos-2 instrument at Gemini South, covering the $1.1 $--$ 1.8~\mu$m range with a spectral resolution of $R \sim 1200$. Standard data reduction procedures were applied, including telluric correction and wavelength calibration. The analysis focused on the identification of emission lines and the estimation of the redshift using cross-correlation techniques and spectral template fitting.
Despite a relatively low signal-to-noise ratio, the spectrum reveals the presence of Pa$\beta$ and Fe\,\textsc{ii} emission lines. The measured redshift is $z = 0.206 \pm 0.001$, which confirms the extragalactic nature of the source. The spectral features such as line ratios and full width at half maximum are consistent with those typically observed in type-1 AGNs, particularly Seyfert 1 galaxies. 
This study demonstrates the ability of near-infrared spectroscopy to reveal AGNs that are obscured by highly extincted and crowded galactic fields. The confirmation of an AGN at low Galactic latitude ($b\sim -6.5$°) shows that near-IR surveys like VVV can successfully penetrate the zone of avoidance. Extending this approach to additional candidates is crucial for improving the census of AGNs hidden behind the Milky Way, as well as for constraining the population of unassociated gamma-ray sources in these troublesome regions.
\end{abstract}




\section{Introduction}
\label{sec:intro}

The Data Release 4 of the Fourth Fermi Large Area Telescope (Fermi-LAT) source catalogue (4FGL-DR4; \citealt{Fermi4FGLDR4}) contains 7194 sources within the energy range of 50 MeV -- 1 TeV. Around 3900 of the identified or associated sources are active galactic nuclei (AGN), and 280 are pulsars. 
Reliable counterparts have yet to be identified for 2460 of the sources, and they are thus known as Unassociated Gamma-ray Sources (UGS). They exhibit significant positional uncertainties, which complicates the identification of counterparts at other wavelengths. 

 The Fermi-LAT catalogues provide the positions of $\gamma$-ray sources, along with the associated uncertainties. These uncertainties are represented by the semi-major and semi-minor axes of an ellipse, as well as the positional angle at the 68\% and 95\% level of confidence.  They are of the order of a few arcminutes in size, a consequence of the limited photon statistics and angular resolution of both the $\gamma$-ray observations and the substantial diffuse $\gamma$-ray emission from the Milky Way (MW). 
 Therefore, UGS represent one of the most significant challenges \citep{Thompson2008} and the key to finding  plausible counterparts is cross-checking with observations at different wavelengths. For instance, observations in radio \citep{Hovatta2014}, infrared  \citep{Raiteri2014} and  sub-millimeter  \citep{LeonTavares2012,LopezCaniego2013} have been previously used. The optical spectroscopic identification of Fermi sources has also been explored 
 \citep{Paggi2014,PenaH2021,GarciaP2023} and X-ray observations carried out with Chandra and Suzaku have proven to be of particular utility in the crowded regions of the Galactic plane \citep{Maeda2011,Cheung2012}. In addition, X-ray data from Swift have been used successfully to confirm the nature of Fermi blazars candidates among the unidentified Fermi-LAT sources \citep{Marchesini2019,Marchesini2020}. Furthermore, the properties of the $\gamma$-ray sources can be used as a statistical set to perform a multivariate analysis \citep{Hassan2013,Doert2014}.

At low Galactic latitudes, most of the unidentified Fermi-LAT sources should have a Galactic origin. The statistical study with non-associated sources of the first Fermi-LAT catalogue \citep{Ackermann2012} showed that a good number of the sources might be blazars. 
\cite{Massaro2011, Massaro2012, Massaro2015} and  \cite{DAbrusco2012, DAbrusco2019} discovered that blazars are well separated from other extragalactic sources in the mid-IR colour space using the Wide-field Infrared Survey Explorer (WISE, \citealt{Wright2010}) data release and identifying the region covered only by blazars.  Although the method is efficient, the selection region is not unique to blazars; some contamination from other types of Active Galactic Nuclei (AGN) or star-forming galaxies remains possible (e.g. \citealt{DAbrusco2014, Kurinsky2017}). Despite this limitation, the approach has proven to be a powerful tool for identifying blazar-like sources within the population of unassociated Fermi-LAT detections.

While blazars constitute the predominant AGN population detected in $\gamma$-rays, the broader class encompasses a diverse array of objects powered by accretion onto supermassive black holes \citep[e.g.,][]{Heckman2014,Netzer2015}. These objects can be classified according to their observational properties, which are influenced by factors such as viewing angle, obscuration, and accretion rate. Among them, type-1 AGN, which are characterized by broad permitted emission lines in their spectra, are of particular relevance.  Their relatively unobscured nature allows for the study of their inner regions and black hole properties. These objects, including Seyfert-1 galaxies and quasars, also exhibit prominent Fe\,\textsc{ii} and hydrogen recombination lines in both the optical and near-infrared (NIR) spectral regimes \citep[e.g.,][]{Joly1981, Osterbrock2006, Riffel2006}. Accurate redshift determination requires broad wavelength coverage to increase the likelihood of detecting multiple emission lines. For certain redshift ranges, only a single strong line may fall within the observed spectral window, which significantly compromises the precision and reliability of the redshift measurement. The presence of several spectral features allows cross-validation between transitions, thereby reducing uncertainties and ensuring a more secure redshift estimate \citep{Ivanov2016,Ivanov2024}.

The VISTA Variables in the Vía Láctea (VVV, \citealt{Minniti2010}) is a deep NIR photometric survey in the $Z$, $Y$, $J$, $H$ and $K_s$ passbands of the Galactic bulge and inner parts of the southern disc. Both the high interstellar extinction and high stellar density make the identification of UGS in these regions even more difficult. \cite{Pichel2020} used the combination of NIR data from VVV and mid-IR data from WISE to study the photometric properties of four known blazars from 
the Roma-BZCAT catalogue \citep{Massaro2015} in these two wavelength regimes. 
Using colour-magnitude and colour-colour diagrams (CMDs and CCDs, respectively), they found for each blazar only one VVV source, coincident with the Roma-BZCAT blazar position, well separated from the rest of the VVV sources found in the error circle of the Fermi-LAT source position. Only two of the four blazars had one WISE source in the blazar region. 

The majority of the surveys have focused on the study of AGN systems at high Galactic latitudes.  However, the census of type-1 AGNs at low Galactic latitudes remains incomplete due to the presence of dust obscuration and source confusion. Recent efforts have begun to identify type-1 AGNs behind the Galactic bulge using NIR observations, where interstellar extinction is significantly lower than at optical wavelengths. \citet{Baravalle2023} studied  several AGN candidates at low Galactic latitudes, combining variability of the sources in the $K_s$ passband  and colour selection criteria using the VVV survey. 
Later, \cite{Donoso2024} applied the method used by \cite{Pichel2020} and \cite{DAbrusco2019} to a sample of UGC sources in the Galactic regions with the goal to identify them with NIR and mid-IR counterparts using the VVV and WISE surveys, respectively. 
With all the sources detected around each Fermi-LAT source, \cite{Donoso2024} made NIR extinction corrected CMDs and CCDs
and found 27 NIR type-1 AGN candidates possibly associated with 14 Fermi-LAT sources using the VVV survey. They also found 2 blazar candidates in the regions of 2 Fermi-LAT sources using WISE data. There was no match between the VVV and WISE candidates. 

These studies reinforce the importance of spectroscopic follow-up in the NIR sources to confirm the AGN nature in regions that have been inaccessible from optical surveys. In this sense, here, we perform a spectroscopic analysis on the potential AGN counterpart. This paper is organized as follows: \S\ref{sec:data} presents the main properties of the AGN candidate; \S\ref{sec:observations} presents the observations and data reduction; the results are discussed in \S\ref{sec:results} and, finally, the main conclusions are in \S\ref{sec:end}.

\section{The counterpart of the unassociated
gamma-ray source 4FGLJ1812.8-3144}
\label{sec:data}

Several attempts were made to confirm the extragalactic nature of the Fermi-LAT unidentified sources using optical spectroscopy.
For instance, \cite{AlvarezC2016} used WISE infrared colours and optical spectra from the Sloan Digital Sky Survey (SDSS, \citealt{York2000}) and the Six-Degree Field Galaxy Survey Database (6dFGS, \citealt{Jones2009}. 
In their sample of Fermi-Large Area Telescope Third Source catalog (3FGL, \citealt{Acero2015}) 
classified as blazar candidates of uncertain type, they found BL Lacs,  radio-loud quasars with flat radio spectrum, and also BL Lacs whose emission is dominated by their host galaxy.  Over the past decade, several groups have also carried out dedicated spectroscopic campaigns, using both archival data and observations from ground-based telescopes, to identify and classify Fermi-LAT sources, notably including the efforts led by \cite{Massaro2015} and more recently by \cite{Lico2022}, among others.
Instead, in the crowded regions of the Galactic bulge and disc, we need to use Flamingos-2 (F2, \citealt{Eikenberry2008, Gomez2012}) NIR spectroscopy to have good quality spectra.  

\cite{Donoso2024} defined the UGC sample taking into account the distributions of positional uncertainties and the interstellar extinction. The A subsample comprises 13 UGS with the lowest Fermi-LAT positional uncertainties, with a semi-major axis of $a$ $<$ 2.5 arcmin, in the MW regions with the lowest interstellar extinctions, A$_{K_s}$ $<$ 1.2 mag. 
In particular, we studied the A12 region associated with the 4FGLJ1812.8-3144 UGS, which has a positional uncertainty of 2.25 arcmin. This source is located in the Galactic bulge, specifically in the b250 VVV tile. 

In this field, three potential counterparts were identified using the NIR CMD and CCD diagrams. Among them, we selected the object VVV-J181258.71-314346.7, located at J2000 coordinates RA  = 18h 12m 58.71s, Dec =  $-31^{\circ} 43' 46.7''$ (l = 0.500$^{\circ}$, b = -6.510$^{\circ}$)  for spectroscopic follow-up. It is the brightest of the three candidates with aperture magnitudes within a radius of 1.02 arcsec of $J = 15.76 \pm 0.03$, $H = 15.09 \pm 0.02$, and $K_s = 14.59 \pm 0.01$ mag obtained from CASU \citep{Emerson2006}.   Also, it shows significant variability, with a $K_s$-band light curve $\sigma_{rms}$ value of 31.6 and a slope of 0.0002 mag day$^{-1}$ over a 1300-day baseline \citep[see][]{Donoso2024}, which is markedly higher than that of the other candidates. Also, it is located closer to the nominal position of the Fermi-LAT source (within 0.7 arcmin) and is situated in a region of low interstellar extinction ($A_{K_s} = 0.14$ mag). Finally, visual inspection of the VVV/VVVX deep-stack images confirms that it displays morphological features consistent with an extragalactic source in the NIR. These combined properties make VVV-J181258.71-314346.7 the most promising AGN candidate and strongly support its selection for spectroscopic observation.

\begin{figure}
\centering
\includegraphics[width=\columnwidth]{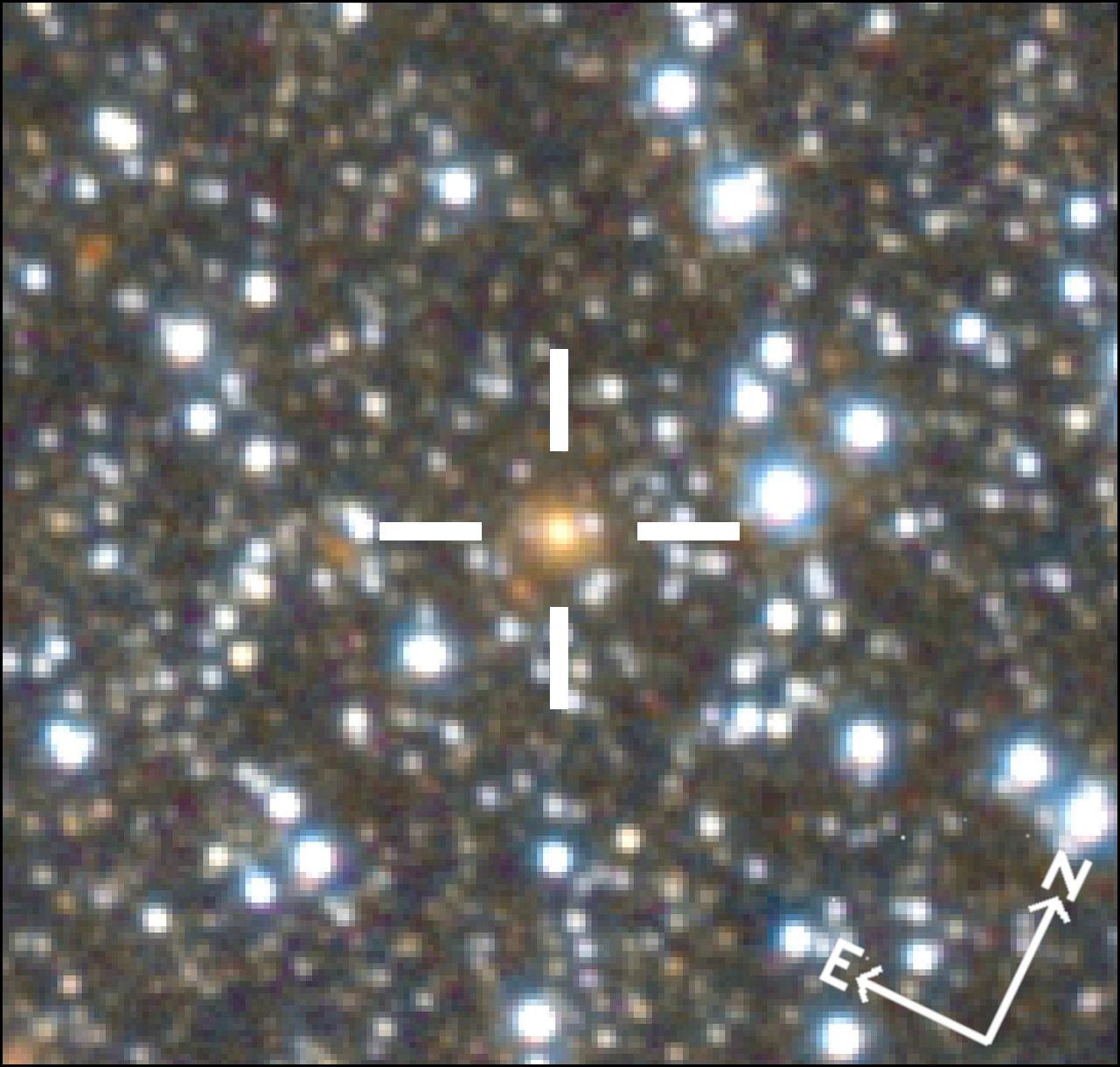}
\caption{$K_s$ (red), $H$ (green), and $J$ (blue) colour composite image of the studied object,  VVV-J181258.71-314346.7. The field-of-view covers 64 arcsec on a side and the orientation is indicated in the bottom-right corner.}
\label{fig:AGN}
\end{figure}

Figure~\ref{fig:AGN} shows a colour composite NIR image of the AGN candidate VVV-J181258.71-314346.7.  This image was obtained using the VVV deep-stack images, created by combining the best-seeing single-epoch $JHK_s$ observations of tile b250 obtained from 2010 to 2023. The typical seeing of these observations is $\sim$ 1 arcsec. The deep stack images for the VVV and the extension, VVVX \citep{Saito2024}, tiles typically reach 1.5 to 2.0 magnitudes deeper in the $K_s$-band, and 0.5 to 0.8 mag deeper in the $J$-band than the individual epochs, revealing fainter details than the single-epoch photometry.

\section{Observations and Data Reduction}
\label{sec:observations}

NIR spectroscopy of the AGN candidate VVV-J181258.71-314346.7 as possible counterpart of the Fermi-LAT unidentified gamma-ray source 4FGLJ1812.8-3144 was performed with the F2 instrument mounted on the 8.1-meter Gemini South telescope at Cerro Pachón, Chile. F2 is equipped with a Hawaii II HgCdTe detector ($2048 \times 2048$ pixels) and offers imaging, long-slit and multi-objects spectroscopy capabilities in the 1.0 -- 2.5~$\mu$m range. Observations were executed in the faint object read mode on April 6, 2022 as part of the GS--2022A--Q--311 program.  They were carried out using the JH grism and the JH filter, which provide coverage from approximately 1.1 to 1.8~$\mu$m. We used the 6-pixel-wide slit, corresponding to 1.08\arcsec, resulting in a spectral resolution of $R \sim 1200$ with a dispersion of $\sim$ 6.5~\text{\AA{}} \text{pixel}$^{-1}$. 
 
We obtained four exposures of 60 seconds each for the science target, using an ABBA nodding pattern to facilitate background subtraction. The observing conditions met the requested constraints, with image quality better than 85\%, low water vapor and cloud cover ($<80$\%), and airmass below 1.5.
To correct for telluric absorption, the star HD-169044 \citep[spectral type F2V,][]{Houk1982} with $H = 8.4$~mag was observed immediately after the science target, using the same instrumental configuration. Four exposures of 5 seconds were acquired for this standard star observed under comparable airmass.

\begin{figure*}
\centering
\includegraphics[width=0.75\textwidth]{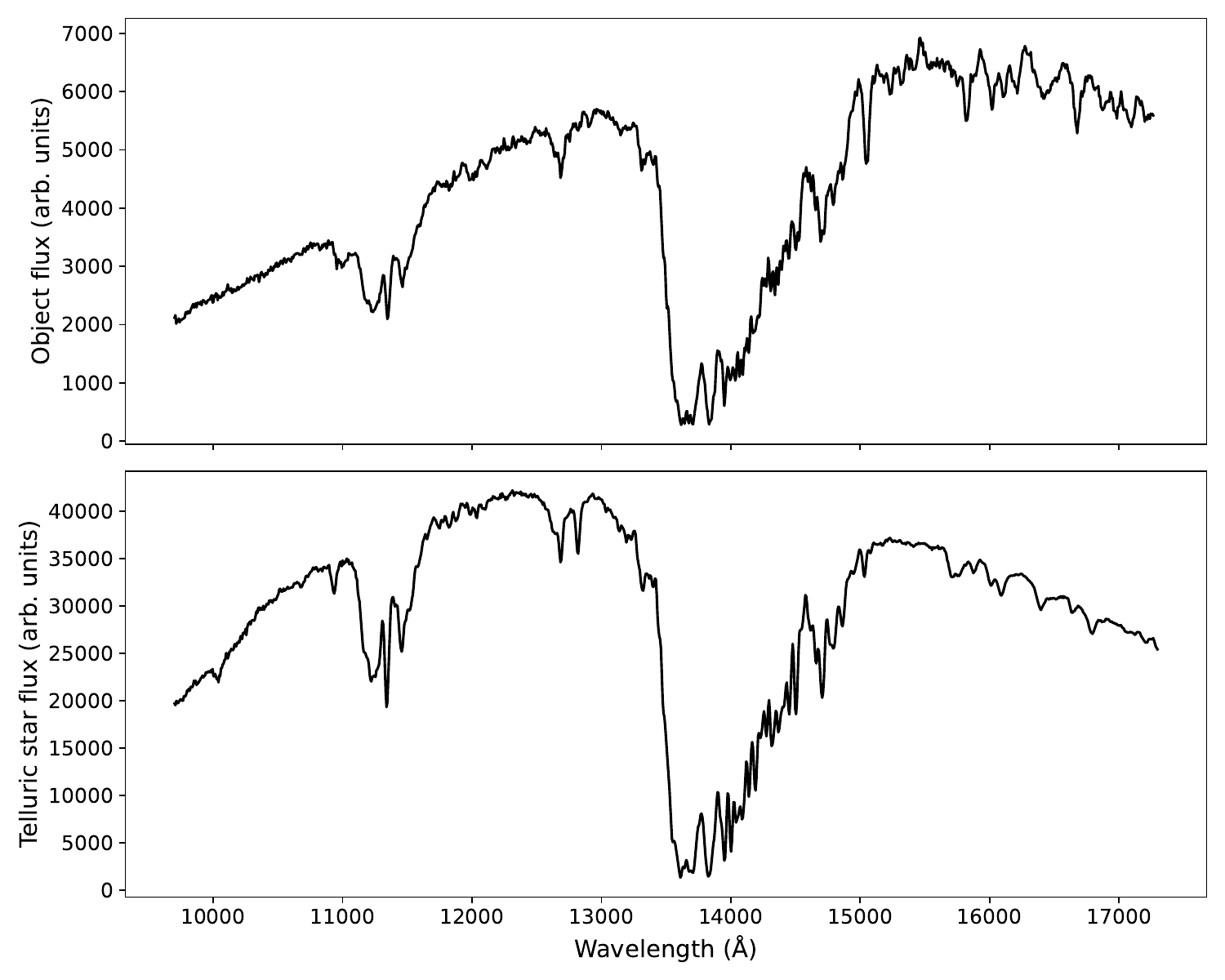}
\caption{The top panel displays the spectrum of the studied object VVV-J181258.71-314346.7, while the bottom panel shows the spectrum of the telluric star HD-169044. Flux is presented in arbitrary units, and wavelength is given in Angstroms.}
\label{fig:espectros_wavelength}
\end{figure*}

Data reduction was carried out using the Gemini IRAF package and the Flamingos-2 PyRAF pipeline. Standard steps included dark subtraction, flat-field correction, wavelength calibration using arc lamp exposures, sky subtraction via the nodded frames, and extraction of one-dimensional spectra. Figure \ref{fig:espectros_wavelength} shows the spectra of the object VVV-J181258.71-314346.7 (top panel) and the telluric star HD-169044 (bottom panel).

Wavelength calibration and telluric correction were performed using standard IRAF tasks. In particular, intrinsic absorption features in the telluric spectrum were modeled and subtracted using Gaussian profiles fitted with the \texttt{splot} task. The resulting spectrum was then divided by a stellar continuum template created using the \texttt{mk1dspec} task in IRAF. The final transmission spectrum obtained this way was used to correct the science spectrum \citep[e.g.,][]{Gaspar2022}. 

\begin{figure*}
\centering
\includegraphics[width=1\textwidth]{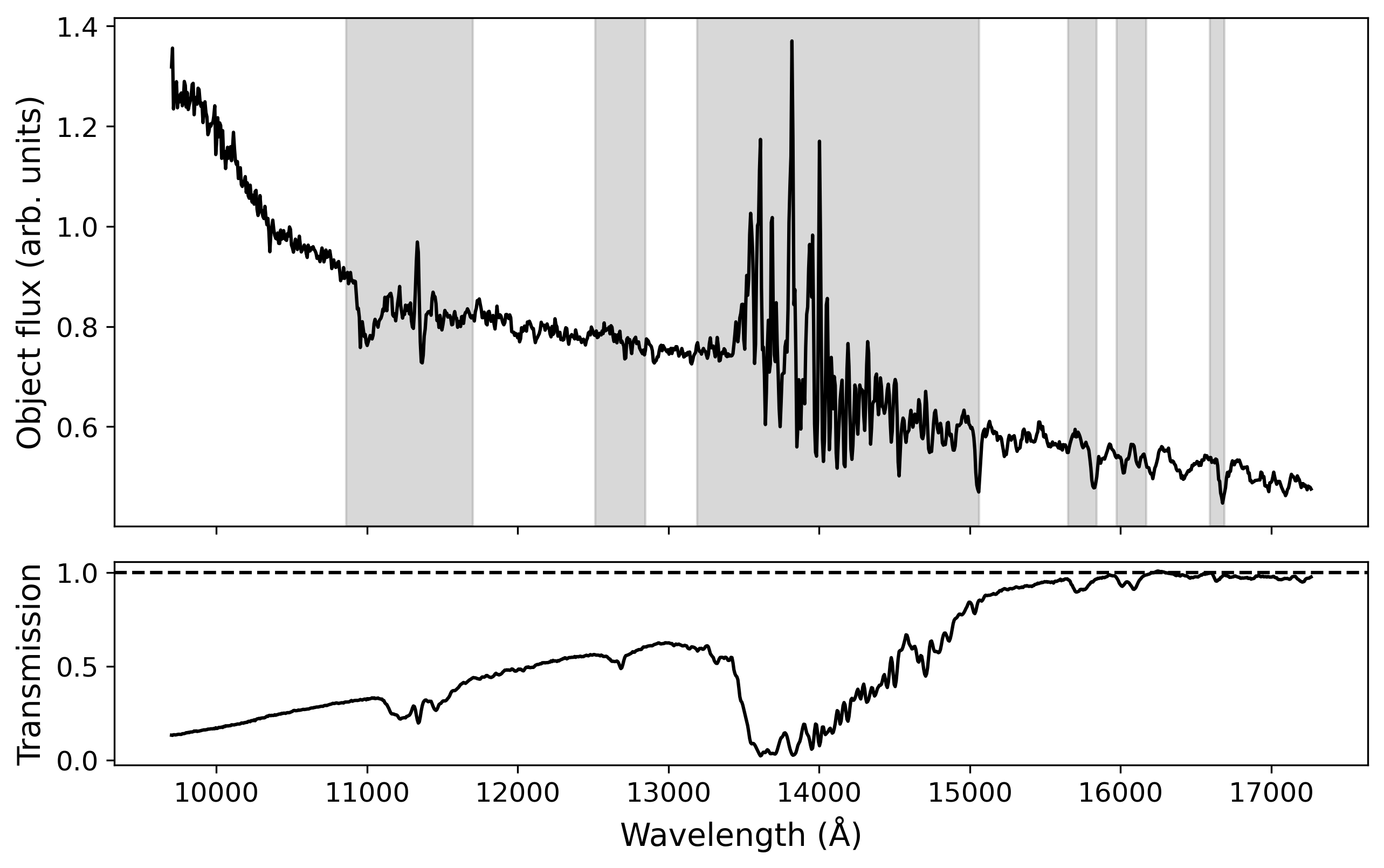}
\caption{Telluric-corrected spectrum of the observed object, VVV-J181258.71-314346.7 (upper panel). The shaded gray regions indicate wavelengths affected by telluric absorption features. Flux is presented in arbitrary units, and wavelength is given in Angstroms. The 
normalized transmission spectrum used for the correction is shown in the lower panel, with a horizontal dashed line at transmission = 1 for reference.}
\label{fig:object}
\end{figure*}

Figure \ref{fig:object} shows the telluric corrected spectrum of the studied object VVV-J181258.71-314346.7 in the upper panel and the transmission spectrum in the lower panel. The resulting NIR spectrum is characterized by low signal-to-noise ratio (S/N), as expected given the observing conditions and the intrinsic faintness of the source. The S/N presents typical values ranging from ~$\sim$4 to 5.5 in the wavelength range of interest, measured as the ratio between the peak flux of each line and the root-mean-square (rms) of the continuum in nearby regions. Although not high enough for detailed analyses, such as kinematic measurements or line ratio diagnostics, it still allows for a reliable redshift estimate based on the detected emission features. Previous studies have demonstrated that robust redshift determinations can be reliably obtained under similar conditions, even in low S/N spectra, when emission lines are identifiable and not severely blended \citep[e.g.,][]{Bolton2012,Masters2019}.

Despite the application of standard telluric correction procedures, several wavelength intervals remain significantly affected by atmospheric absorption features, particularly in regions of strong and variable water vapor opacity. These residuals can introduce uncertainties in the local continuum level and may hinder the reliable identification of faint or blended features. For clarity and to guide the interpretation of the spectra, we explicitly mark these regions in grey in the figure. Caution should be exercised when evaluating any spectral features within these shaded areas, as their interpretation may be compromised by incomplete or imperfect telluric subtraction.

\section{The spectroscopic redshift}
\label{sec:results}

It is important to note, despite the modest spectral S/N, that a preliminary visual inspection of both the wavelength-calibrated spectrum, prior to the implementation of telluric correction, and the final telluric-corrected spectrum revealed the presence of at least two prominent emission features.  The most prominent lines manifest at approximately 15467~\AA{} and 15932~\AA{}, which are interpreted as red-shifted emission lines.

The estimation of the redshift of the source was accomplished through the construction of a synthetic spectral template using the IRAF task \texttt{mk1dspec}. The template under consideration includes three emission lines that are commonly observed in the NIR spectra of both AGNs and star-forming galaxies. These lines include the hydrogen recombination line Pa$\beta$ at 12822~\AA{}, as well as two Fe\,\textsc{ii} emission features at 12570~\AA{} and 13209~\AA{} \citep[see, e.g.,][]{Riffel2006, Schonell2025}. The selection of these lines was based on two fundamental considerations. Firstly, their prominence in type-1 AGNs, particularly in Seyfert galaxies, where Fe\,\textsc{ii} emission is prevalent. Secondly, their potential appearance in star-forming systems, where Fe\,\textsc{ii} may be excited by shocks or outflows. 
Given the limiting magnitudes of the VVV survey, it is unlikely to detect compact galaxies at redshifts significantly beyond $z \sim 0.25$. This is supported by previous spectroscopic studies of VVV sources. For example, \citet{Baravalle2019} measured redshifts for galaxies in the VVV-J144321.06–611753.9 cluster, identifying a mean cluster redshift of $z = 0.234 \pm 0.022$ from the two brightest members ($K_s = 13.7$ and 14.8 mag). Similarly, \citet{Galdeano2023} obtained redshifts for five galaxies in the VVVGCl-B-J181435-381432 cluster, with total $K_s$ magnitudes between 13.69 and 15.18 mag and a mean redshift of $z = 0.225 \pm 0.014$. These results indicate that VVV is sensitive to galaxies up to $z \sim 0.25$, particularly for moderately bright sources like ours. Therefore, our template was tailored to identify emission lines that would lie within the observed wavelength range for $z \lesssim 0.25$.

Subsequently,  the \texttt{fxcor} task from the IRAF RV package was used to execute the cross-correlation between the observed spectrum of the studied source and the synthetic template in Fourier space.   This procedure was carried out to identify the velocity shift that optimizes the correlation between spectral features \citep{TonryDavis1979}. Its primary function is to measure radial velocities in cases where data are characterized by noise or low resolution, particularly when individual features are obscured, indistinct, or weak. In our case, the cross-correlation yielded a radial velocity of $\sim$ 61864~km~s$^{-1}$, corresponding to a redshift of $z \sim 0.206$. The quality of the correlation is quantified by the Tonry \& Davis R-value, which measures the ratio of the peak height of the correlation to the average noise level in the Fourier domain. Values above $\sim5$ are generally considered to provide a secure radial velocity estimate, particularly in low S/N regimes. We measured an R-value of 6.26, indicating a reasonably reliable measurement.

\begin{figure*}
\centering
\includegraphics[width=1\textwidth]{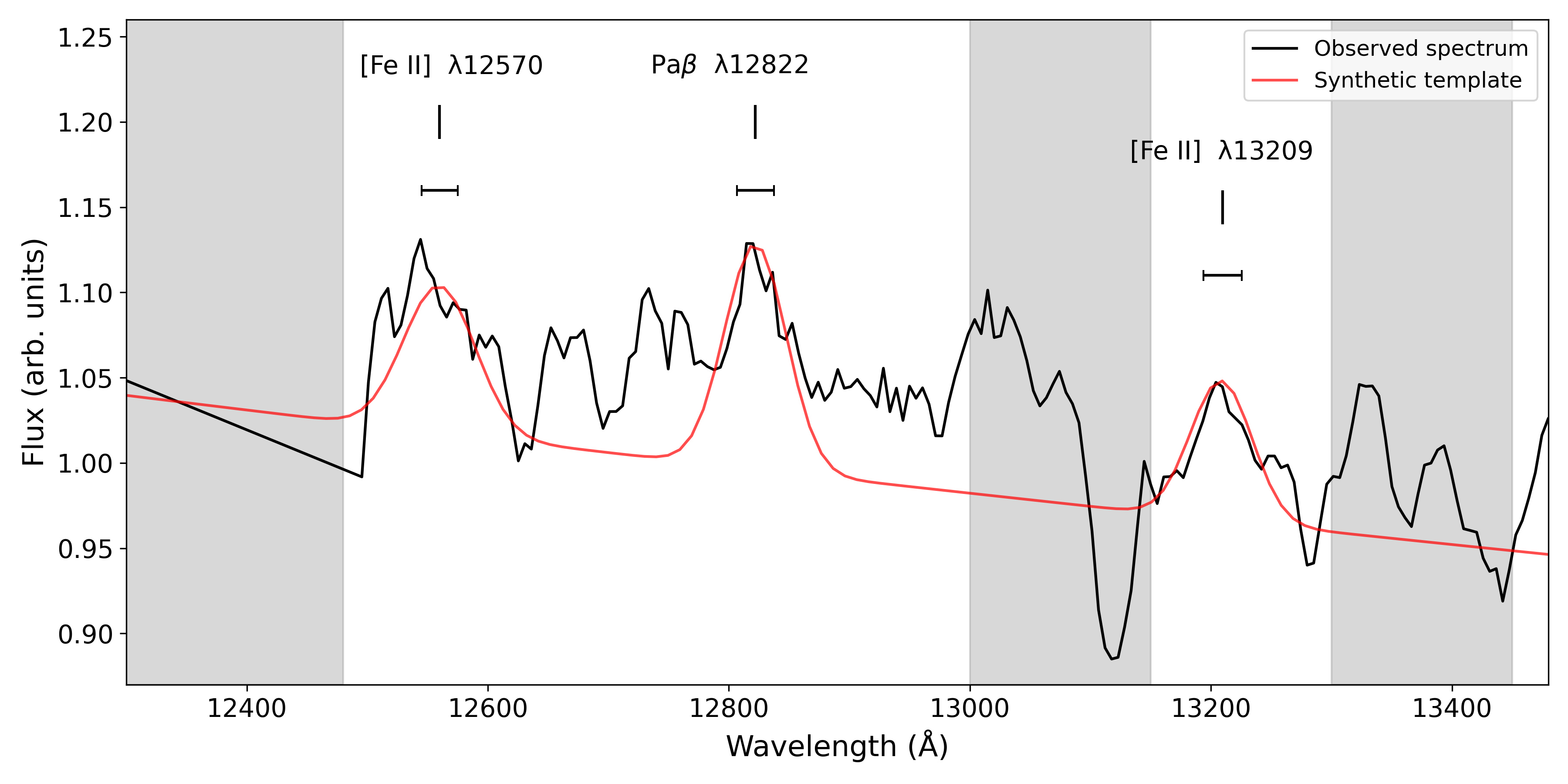}
\caption{Zoomed rest-frame spectrum considering the calculated redshift of $z = 0.206$. The black curve shows the observed telluric-corrected spectrum, while the red curve corresponds to the synthetic template used for the cross-correlation. Vertical lines indicate the central wavelengths of the identified emission lines (Fe\,\textsc{ii}$\lambda12570$, Pa$\beta\lambda12822$, and Fe\,\textsc{ii}$\lambda13209$) in the rest frame. Horizontal error bars represent the uncertainty in the redshift determination ($\pm 0.001$). The shaded gray regions indicate wavelengths affected by telluric absorption features and are the same as in the previous figure (see Sect.~\ref{sec:observations}). Flux is in arbitrary units and wavelength is given in Angstroms.}
\label{fig:lineas}
\end{figure*}

In order to calculate the redshift uncertainty, we complemented the cross-correlation result with individual Gaussian fits to the three detected emission lines: Fe\textsc{ii}$\lambda$12570, Pa$\beta$$\lambda$12822, and Fe\textsc{ii}$\lambda$13209. Each line was independently fitted multiple times, resulting in individual redshift measurements. These values were then combined to evaluate the statistical dispersion among the lines, which we adopted as a proxy for the redshift uncertainty. The resulting 1$\sigma$ error is 0.00066, which we conservatively round to 0.001. We therefore report a redshift of $z = 0.206 \pm 0.001$, where the uncertainty reflects both the scatter among the line-based measurements and the limitations imposed by the spectral resolution and the signal-to-noise ratio. Figure~\ref{fig:lineas} shows a zoomed-in view of the rest-frame spectrum based on the derived redshift, where the previously described synthetic template is overplotted, vertical lines indicate the expected positions of the emission features at their corresponding wavelengths, and horizontal error bars illustrate the redshift uncertainty of $\pm 0.001$.

\section{Line ratios and full width at half maximum}
\label{sec:fwhm}

The relative intensities of the detected emission lines offer valuable information about the physical conditions of the emitting gas and the excitation mechanisms involved. A particularly useful diagnostic is the ratio between Fe\,\textsc{ii}$\lambda12570$ and Pa$\beta$, which traces the balance between collisional excitation and recombination processes in the broad-line region \citep[e.g.,][]{Mouri1993, Goodrich1994, Draine2011, Terao2016}. The Fe\,\textsc{ii} emission is expected to originate in partially ionized zones with high column density and significant microturbulence, typically associated with nuclear activity and accretion \citep[e.g.,][]{Netzer1983, Forbes1993, Verner2003, Baldwin2004}. 

In Seyfert galaxies, values of the ratio Fe\,\textsc{ii}$\lambda12570$/Pa$\beta$ usually range between 0.5 and 4.6 \citep[e.g.,][]{Alonso-Herrero1997,Rodriguez-Ardila2004, Ramos-Almeida2009, Terao2016}. For our source, we obtain Fe\,\textsc{ii}$\lambda12570$/Pa$\beta = 2.6 \pm 1.3$, which, despite the relatively high uncertainty due to the modest signal-to-noise ratio, lies well within the expected interval for AGN. Similarly, the ratio Fe\,\textsc{ii}$\lambda12570$/Fe\,\textsc{ii}$\lambda13209 = 1.4 \pm 0.6$ is consistent with measurements reported in other active nuclei \citep[e.g.,][]{Riffel2006, Ramos-Almeida2009}, further supporting the interpretation that the detected Fe\,\textsc{ii} features are powered by nuclear activity rather than by stellar processes.  

We also measured the observed full widths at half maximum (FWHM) of the three detected emission lines to evaluate whether the line broadening is intrinsic or dominated by instrumental effects. The Fe\textsc{ii}$\lambda$12570 and Fe\textsc{ii}$\lambda$13209 lines exhibit FWHM$_{\rm obs}$ values of $104 \pm 9$~\AA{} and $105 \pm 12$~\AA{}, respectively, while the Pa$\beta$ line at 12822~\AA{} shows $FWHM_{\rm obs} = 70 \pm 7$~\AA{}. The instrumental resolution of our data, with $R \sim 1200$, corresponds to $FWHM_{\rm inst} \sim 12.5$~\AA{} at 15000~\AA{} (i.e., $\sim 250$~km~s$^{-1}$). Correcting for instrumental broadening as $FWHM = \sqrt{FWHM_{\rm obs}^2 - FWHM_{\rm inst}^2}$, we derive $FWHM = 103 \pm 9$~\AA{} and $FWHM = 104 \pm 12$~\AA{} for the Fe\,\textsc{ii} lines, respectively and $69 \pm 7$~\AA{} for Pa$\beta$, corresponding to velocity dispersions of $\sim 2400 \pm 250$~km~s$^{-1}$ and $\sim 1600 \pm 150$~km~s$^{-1}$, respectively. Although the low S/N ratio of the spectra introduces significant uncertainties in these measurements, the derived FWHM values are substantially larger than the instrumental resolution. This suggests the presence of intrinsically broadened emission lines, consistent with high-velocity gas in the broad-line region, supporting the classification of VVV-J181258.71-314346.7 as a type-1 AGN \citep[e.g.,][]{Riffel2006, Ramos-Almeida2009, denBrok2022}.

Taken together, the emission-line ratios and the line widths are consistent with the values typically reported for type-1 AGNs, reinforcing the classification of VVV-J181258.71-314346.7 as an active galactic nucleus of this type.

\section{Conclusions and final comments}
\label{sec:end}

In this work, we present the spectroscopic study of a candidate counterpart to the unassociated gamma-ray source 4FGLJ1812.8-3144, the VVV-J181258.71-314346.7. The source was selected based on its NIR photometry, its location in the NIR colour-colour diagrams, and its moderate variability in the $K_s$ light curves.  We have obtained spectroscopic data using F2 at Gemini South. Although the resulting spectrum exhibits a relatively low signal-to-noise ratio, we were able to identify prominent emission lines corresponding to Pa$\beta$ and two Fe\,\textsc{ii} transitions, from which we derived a redshift of $z = 0.206 \pm 0.001$. The redshift estimate is based on these three distinct emission lines, providing additional robustness to our measurement.

The VVV-J181258.71-314346.7 source had previously classified  as a type-1 AGN candidate based on its NIR photometric properties and variability \citep{Donoso2024}. The result presented here confirms the extragalactic nature of this source. Our spectroscopic detection of Pa$\beta$ and Fe\,\textsc{ii} emission strongly supports this classification, as these features are typical of Seyfert-1 galaxies.  These galaxies are well known to exhibit Fe\,\textsc{ii} emission over a wide range of wavelengths, both in optical and NIR wavelengths. In the optical, Fe\,\textsc{ii} lines are among the most characteristic features of type-1 AGNs, as extensively documented in the literature \citep[e.g.,][]{Phillips1978FeII,Joly1981,Vestergaard2005FeII, Schmidt2018,Schmidt2021}. NIR Fe\,\textsc{ii} emission has also been widely reported in Seyfert-1 and narrow-line Seyfert-1 galaxies, particularly in the $z \sim 0.1$ -- 0.3 range where the relevant transitions fall in the $H$ band \citep[e.g.,][]{Moorwood1988FeII, RodriguezArdila2000FeII, RodriguezArdila2002FeII,Reunanen2002FeII}. The spectral properties of VVV-J181258.71-314346.7 are fully consistent with these trends, reinforcing its identification as a type-1 AGN, most likely a Seyfert 1 galaxy.

In addition, the analysis of the relative intensities of the detected emission lines and their intrinsic widths further support the AGN nature of the source. The measured Fe\textsc{ii}$\lambda12570$/Pa$\beta$ ratio of $2.6 \pm 1.3$ falls within the range typically observed in Seyfert galaxies, while the Fe\textsc{ii}$\lambda12570$/Fe\textsc{ii}$\lambda13209$ ratio of $1.4 \pm 0.6$ is also consistent with values reported in active nuclei. These line ratios suggest that Pa$\beta$ and the Fe \textsc{ii} emission are powered by nuclear activity rather than by star formation processes. Moreover, the intrinsic FWHM values derived for the Fe \textsc{ii} and Pa$\beta$ lines (in the range of $\sim1600$ to $2400$ km s$^{-1}$) are significantly broader than the instrumental profile, confirming the presence of high-velocity gas in the central regions. Together, these diagnostics provide compelling evidence that VVV-J181258.71-314346.7 is a type-1 AGN, most likely a Seyfert galaxy. Although the spectral properties of VVV-J181258.71-314346.7 are consistent with a classical Seyfert 1 classification, the modest signal-to-noise ratio of the spectrum prevents a definitive exclusion of a narrow line Seyfert 1 (NLS1) nature. While NLS1 galaxies are more frequently associated with gamma-ray emission \citep[e.g.,][]{Foschini2015, Paliya2019}, such emission has also been observed in classical Seyfert 1 galaxies \citep[e.g.,][]{Wojaczynski2015, Michiyama2024}. Therefore, the identification of VVV-J181258.71-314346.7 as the counterpart to 4FGLJ1812.8-3144 is consistent with its observed spectral properties and supported by previous reports of gamma-ray emission in classical Seyfert 1 galaxies.

These results highlight the potential of NIR spectroscopy to unveil obscured AGNs in the most challenging regions of the sky, such as the Galactic bulge and disc. Expanding this type of analysis to other promising candidates in the VVV survey will be key to improving our census of AGNs and refining the population statistics of gamma-ray sources at low Galactic latitudes.

\section*{Acknowledgements}
We want to thank our referee for constructive comments and suggestions that improved this work. We are also grateful to G. Ferrero for fruitful discussions. The authors thank R. Vena Valdarenas for helping us improve some figures.
This research was partially supported by Consejo de Investigaciones Cient\'ificas y T\'ecnicas (CONICET) and Secretar\'ia de Ciencia y T\'ecnica de la Universidad Nacional de C\'ordoba (SeCyT). 
Based on observations obtained at the international Gemini Observatory, a program of NSF NOIRLab, which is managed by the Association of Universities for Research in Astronomy (AURA) under a cooperative agreement with the U.S. National Science Foundation on behalf of the Gemini Observatory partnership: the U.S. National Science Foundation (United States), National Research Council (Canada), Agencia Nacional de Investigación y Desarrollo (Chile), Ministerio de Ciencia, Tecnología e Innovación (Argentina), Ministério da Ciência, Tecnologia, Inovações e Comunicações (Brazil), and Korea Astronomy and Space Science Institute (Republic of Korea).
We made use of the cross-match service provided by CDS, Strasbourg.
The authors gratefully acknowledge data from the ESO Public Survey program IDs 179.B-2002 and 198.B-2004 taken with the VISTA telescope, and products from the Cambridge Astronomical Survey Unit (CASU). We also thank the team responsible for the deep stacks project, an advanced product of the VVV/VVVX survey led by N. Cross and M. Read, carried out at the Wide Field Astronomical Unit (WFAU) of the Royal Observatory of Edinburgh (ROE). 

\section*{Data Availability}

The data underlying this article will be shared on reasonable request to the corresponding author. 



\bibliographystyle{mnras}
\bibliography{Bibliography} 








\bsp	
\label{lastpage}
\end{document}